\documentclass{aa}
\usepackage{graphicx}

\def\gsim{\vcenter{\hbox{$>$}\offinterlineskip\hbox{$\sim$}}}

\begin{document}

\title{No disks around low-mass stars and brown dwarfs in the young $\sigma$\,Orionis cluster?}

\author{J.M. Oliveira\inst{1},
   R.D. Jeffries\inst{1}, 
   M.J. Kenyon\inst{1}, 
   S.A. Thompson\inst{1} 
   \and Tim Naylor\inst{1,2}}

\offprints{joana@astro.keele.ac.uk}

\institute{Department of Physics, Keele University, Staffordshire ST5 5BG,  United Kingdom \and
School of Physics, University of Exeter, Exeter EX4 4QL, United Kingdom }

\date{Received ; accepted }

\authorrunning{Oliveira et al.}

\abstract{We report on the analysis of 2MASS near-infrared data of a sample of low-mass stars and brown dwarfs in the $\sigma$\,Orionis cluster. Youth and cluster membership have been spectroscopically confirmed using the Li~{\sc i} spectral line. We find little evidence in the $JHK_{\rm s}$ colour-colour diagram for near-infrared excess emission for these cluster members. By comparison with model expectations, at most 2 out of 34 stars show $(H-K_{\rm s})$ colour consistent with a near-infrared excess. This scarcity of near-infrared signatures of circumstellar disks in the lower-mass and substellar regimes of this cluster contrasts with findings in younger clusters, hinting at an age dependence of the disk frequency. Taking into account the apparent cluster age, our result supports the idea of a relatively fast (few Myr) disk dissipation and extends this conclusion to the substellar regime. We also find some evidence that, in this cluster, the disk frequency as measured by the K$_{\rm s}$-band excess may be mass dependent.
}

\maketitle

\section{Introduction}

Star formation is believed to start with the fragmentation of rotating and magnetized clouds of interstellar gas, and further gravitational collapse forms stars across the whole mass range. Flattened disk-like structures arise during this star formation process (see review by Evans \cite{evans99}), and circumstellar disks are common around young stars. Even though it is thought that brown dwarfs are just the substellar counterparts of stars, forming as outlined above, it has also been proposed that brown dwarfs form either through instabilities in circumstellar disks (e.g.\ Pickett et al.\ \cite{pickett00}) or as stellar embryos ejected from multiple proto-stellar systems (Reipurth \& Clarke \cite{reipurth01}).

Disk dissipation timescales, together with the environment of the young stellar population, are of particular importance, in the context of pre-main-sequence stellar evolution and of planet formation processes. Haisch et al.\ (\cite{haisch01}) have found evidence for an age dependence of the disk frequency from the analysis of low-mass stars in several young clusters in OB associations and star forming regions. Even though the importance of the cluster environment is not yet clear, it has been proposed that photo-evaporation by the massive OB stars plays a role in disk dissipation (e.g.\ Johnstone et al. \cite{johnstone98}; Bally et al.\ \cite{bally00}). It is also relevant to probe how the observed disk properties extend into the substellar regime.

\object{$\sigma$\,Orionis} is a member of the Orion\,OB1b association, with an estimated age of 1.7$-$7\,Myr and a distance modulus of 7.8$-$8\,mag (e.g.\ Warren \& Hesser \cite{warren78}; Brown et al.\ \cite{brown94}) --- the Hipparcos distance to this star is 352$^{+166}_{-85}$\,pc. ROSAT observations and follow-up photometry (Wolk \cite{wolk96}) revealed a cluster of low-mass young stellar objects near $\sigma$\,Ori. The reddening towards this massive O-type star is low ($E(B-V)$\,=\,0.05\,mag, e.g.\ Brown et al.\ \cite{brown94}), suggesting that the young stellar population is affected by low extinction.

Kenyon et al.\ (\cite{kenyon01}) have confirmed both youth and cluster membership for a sample of low-mass and substellar objects within 30\arcmin\ of $\sigma$\,Ori. We analyse 2MASS near-infrared photometric data for their sample of stars. The $JHK_{\rm s}$ colour-colour diagram of these cluster members shows little evidence for near-infrared excess. By comparing observed $(H-K_{\rm s})$ with model expectations, we find that only 2 objects out of 34 show an infrared excess suggestive of circumstellar disks. This is in striking contrast to high disk frequencies determined for the low-mass populations in younger clusters (Trapezium, NGC\,2024 and NGC\,2264, Haisch et al.\ (\cite{haisch01}) and references therein). Our result provides valuable support for the age dependence of disk frequency in low-mass stars and may indicate that this dependence continues into the substellar regime. 


\section{Target selection and near-infrared data}

An R- and I-band survey, obtained with the Wide-Field Camera (WFC) at the 2.5\,m Isaac Newton Telescope (INT) and covering approximately 1 square degree centered on $\sigma$\,Ori, was used by Kenyon et al.\ (\cite{kenyon01}) to select low-mass and brown dwarf cluster candidates for follow-up optical spectroscopy. These photometric candidates were selected as stars younger than about 100\,My, according to Baraffe et al.\ (\cite{baraffe98}) isochrones. High-resolution spectroscopy in the 6400$-$6800\,\AA\ range was obtained at the 4.2\,m William Herschel Telescope (WHT), with the Wide-Field Fiber Optic Spectrograph (WYFFOS). From the 83 spectroscopically observed candidates: 39 are lithium rich, based on their Li\,{\sc i} ($\lambda$\,6707\,\AA) equivalent widths, confirming their status as very young low-mass stars and brown dwarfs and almost certainly cluster members; 44 objects show no evidence for lithium and are therefore likely (though see below) to be non-members. This sample of stars is described in detail elsewhere (Kenyon et al.\ \cite{kenyon01}; Kenyon et al.\ in preparation).

We have searched the Point Source Catalog of the 2MASS Second Incremental Data Release for infrared counterparts of this sample of cluster members. Information on the 2MASS survey and catalogue products can be found in Beichman et al.\ (\cite{beichman98}) and Cutri et al.\ (\cite{cutri01}), respectively. Of the 39 sample objects, 34 have 2MASS counterparts within a radius of 0.4$\arcsec$ of the optical sources (of the 44 lithium-poor objects, 43 have 2MASS counterparts).These stars were detected in the J-, H- and K$_{\rm s}$-bands and the observations are not contaminated by artifacts or confusion with bright or extended sources.

We want to infer the disk fraction amongst the 34 cluster members. We analyse the $JHK_{\rm s}$ colour-colour diagram of the cluster members, together with a control sample of photometric cluster candidates: 43 lithium-poor sources (see above) and 25 photometric candidates from B\'{e}jar et al.\ (\cite{bejar01} and references therein). The role of the control sample is to assess whether or not the spectroscopic selection method is biased against stars with disks. Photospheric spectral lines in classical T Tauri stars can be heavily veiled at optical (e.g.\ Gullbring et al.\ \cite{gullbring98}) and infrared (Folha \& Emerson \cite{folha99}) wavelengths. In particular, the hot continuum attributed to disk accretion can fill-in the Li\,{\sc i} 6707\,\AA\ line (e.g.\ Magazz\`{u} et al.\ \cite{magazzu92}). If the colour-colour diagram of the control sample revealed a significant fraction of objects with near-infrared excess, then the spectroscopically confirmed sample could be biased.


\section{Results}

\begin{figure}[t]
\resizebox{\hsize}{!}{\includegraphics{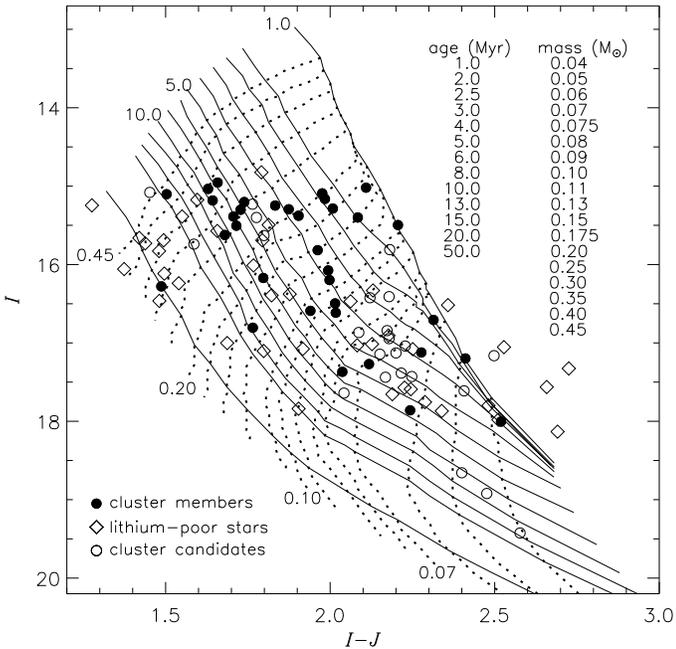}}
\caption{$I$/($I-J$) colour-magnitude diagram for the cluster members and the control sample (see text). The Baraffe et al.\ (\cite{baraffe98}) evolutionary tracks (dotted line) and isochrones (full line) were used to estimate the masses and ages of these objects. A Hipparcos distance to the cluster $d$\,=\,352\,pc and optical extinction $A_{\rm V}$\,=\,0.15\,mag are adopted.}
\label{f1}
\end{figure}

We have determined masses and ages for the cluster members using the $I$/($I-J$) colour-magnitude diagram and Baraffe et al.\ (\cite{baraffe98}) evolutionary models (Fig.\,\ref{f1}). The sample populates the mass range of 0.04$-$0.45\,M$_{\odot}$ and the age range of 1$-$50\,Myr. The median age of the sample is 4.2$^{+2.7}_{-1.5}$\,Myr, where the uncertainty is a very conservative estimate based on the error in the Hipparcos distance to $\sigma$ Ori. If the objects in the control sample {\em are} cluster members, then they have similar masses and ages (Fig.\,\ref{f1}). Of the 34 cluster members, 7 have masses below the hydrogen burning limit ($M\sim0.08$\,M$_{\odot}$) and are candidate brown dwarfs (5$-$9 brown dwarf candidates if the extreme distance uncertainties are taken into account). From Fig.\,\ref{f1}, 4 objects seem to be older than about 10\,Myr and therefore there are some doubts about their cluster membership. The observed age spread could either be real, in the sense that star formation in this cluster lasted for a few million years, or it could just be a combination of binarity, photometric uncertainties and source variability (the I- and J-band observations are non-simultaneous).

\begin{figure*}
\centering
\includegraphics[width=17cm]{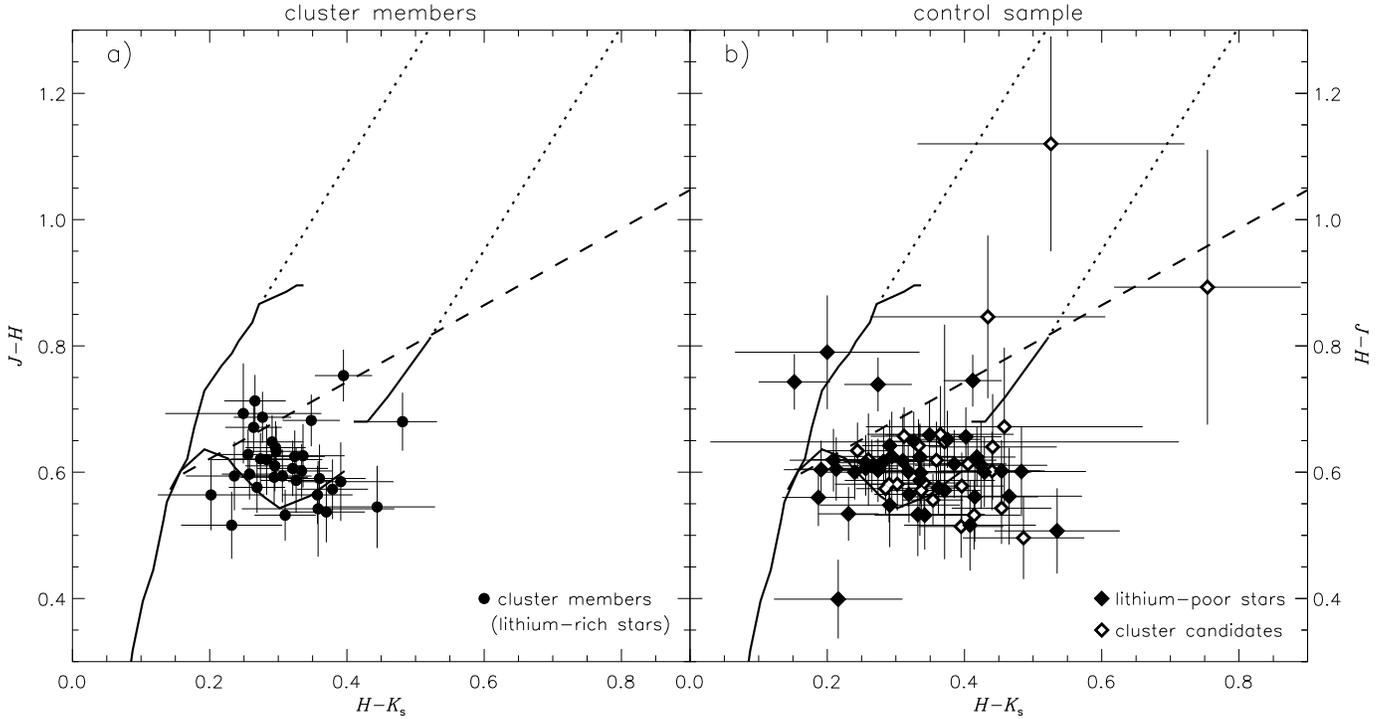}
\caption{$JHK_{\rm s}$ colour-colour diagram for a) the cluster members and b) the control sample (see text). The loci of main-sequence and giants stars (Bessell \& Brett \cite{bessell88}) and of M8.5$-$M9.5 dwarfs (Kirkpatrick et al.\ \cite{kirkpatrick00}) are plotted (full line), as well as the locus of classical T Tauri stars (dashed line, Meyer et al.\ \cite{meyer97}). Reddening bands for the giant stars and for late-M dwarfs are also plotted (dotted lines). The cluster members show no significant evidence for near-infrared excess.}
\label{f2}
\end{figure*}

In Fig.\,\ref{f2} we show the $JHK_{\rm s}$ colour-colour diagram for the sample of cluster members (a) and for our control sample (b). We plot the empirical loci of main-sequence and giant stars (Bessell \& Brett \cite{bessell88}), M8$-$M9.5 dwarfs (Kirkpatrick et al.\ \cite{kirkpatrick00}) and classical T Tauri stars (Meyer et al.\ \cite{meyer97}). All these loci were converted to the 2MASS photometric system using the colour-colour transformations from Carpenter (\cite{carpenter01}). Also in this figure are relevant reddening bands (Cohen et al.\ \cite{cohen81}).

From Fig.\,\ref{f2}a no cluster member shows unequivocal near-infrared excess, as none of these stars appear significantly to the right of the reddening band for M dwarfs. From the 68 objects in the control sample, at most one lithium-poor star and three photometric candidates could be said to possess near-infrared excess (Fig.\,\ref{f2}b), albeit with little confidence (less than a 2$\sigma$ ($H-K_{\rm s}$) excess detection). This demonstrates that there is no significant population of heavily veiled, low-mass T Tauri objects with weak Li\,{\sc i} 6707\,\AA\ lines, but large near-infrared excesses.

From Fig.\,\ref{f2}a alone, it is not possible to ascertain whether a low-reddening object with $(H-K_{\rm s})$ $\sim$ 0.4$-$0.5\,mag is a low-mass star with purely photospheric H- and K$_{\rm s}$-band magnitudes or a slightly more massive star with considerable $(H-K_{\rm s})$ excess. To do this we have parameterised possible excesses by using $\Delta(H-K_{\rm s}) = (H-K_{\rm s}) - (H-K_{\rm s})_{0}$, where $(H-K_{\rm s})_{0}$ are the Baraffe et al. (\cite{baraffe98}) predictions for photospheric $(H-K_{\rm s})$ colours according to the ages and masses derived from Fig.\,\ref{f1}. When compared with the photometric uncertainties, we find that only one cluster member exhibits significant $(H-K_{\rm s})$ excess ($\sim$ 3$\sigma$ detection) and another object just a 1$\sigma$ detection. Both objects have masses in the range 0.15$-$0.175\,M$_{\odot}$. The near-infrared colours from the Baraffe models seem to be slightly redder than the observed colours (as also noted by Lucas et al. \cite{lucas01}). Even taking this into account, at most two stars show a significant $(H-K_{\rm s})$ excess. 


\section{Discussion and conclusions}

Our sample of low-mass $\sigma$\,Ori cluster members shows an extremely low fraction of objects with near-infrared excess: at most 2 out of 34. As near-infrared excesses in pre-main-sequence stars are normally attributed to circumstellar disks, we interpret this result as a 6\% disk frequency in the $\sigma$\,Ori cluster (7\% if the 4 apparently older stars are removed from the sample as possible contaminants). Using an X-ray selected sample ($M \gsim$ 0.5\,M$_{\odot}$), Wolk (\cite{wolk96}) proposed a 12\% (6/49) disk frequency, using H$\alpha$ emission as an indirect indicator of circumstellar material (emission equivalent width $>$\,10\,\AA). X-ray selection might bias the sample against stars with disks (Stelzer \& Neuh\"{a}user \cite{stelzer01}). However, H$\alpha$ samples are biased towards stars with high accretion rates. Stassun et al.\ (\cite{stassun99}) show that some low-mass stars (in Orion) with K-band excess do not show strong H$\alpha$ emission. For both these reasons, Wolk's estimated disk fraction must be a lower limit to the fraction one would estimate from near-infrared photometry. Therefore, there is some evidence, tempered by small number statistics, that the disk frequency (as measured by K-band excess) decreases towards lower masses in the $\sigma$\,Ori association.

As the near-infrared colours trace the warmer dust, the magnitude of the near-infrared excess depends on the position and temperature of the inner disk boundary. Towards very low masses, the stellar effective temperature is critical for the disk detection efficiency at these wavelengths. Natta \& Testi (\cite{natta01}) show that classical T Tauri disk models describe well the observed spectral energy distribution (SED) for brown dwarf candidates. They suggest that, in the substellar regime, the inner parts of such disks might not be hot enough to produce detectable near-infrared excess. Indeed, Comer\'{o}n et al.\ (\cite{comeron00}) find that several low-mass stars and brown dwarfs (0.03\,M$_{\odot}\leq M\leq$\,0.2\,M$_{\odot}$) in Chamaeleon\,I exhibit 6.7\,$\mu$m excess emission but no K-band excess. It thus seems possible that towards less massive objects one would expect to find comparatively fewer objects with near-infrared excess and perhaps none among brown dwarfs. However, Muench et al.\ (\cite{muench01}) find a K$_{\rm s}$-band excess in 65\% of Trapezium cluster brown dwarfs, slightly more than the 50\% K-band excess fraction that Lada et al.\ (\cite{lada00}) find in more massive stars (0.1$-$1\,M$_{\odot}$). 

Clearly, the K-band excess frequency and presumably the fraction of low-mass stars and brown dwarfs possessing warm circumstellar material is smaller around $\sigma$\,Ori than in the 1\,Myr old Trapezium cluster. This is true for both the low-mass stars and also for the brown dwarfs. Even though our sample of brown dwarfs is small, the lack of even a single K$_{\rm s}$-band excess detection (in the $JHK_{\rm s}$ colour-colour diagram) is significant compared with the 65\% fraction seen by Muench et al. (\cite{muench01}) in the Trapezium. Our results are consistent with the low disk fractions of 5$-$12\%, determined from K- and L-band excesses and H$\alpha$ emission in the similarly aged Upper Scorpius, NGC\,2362 and $\lambda$\,Ori associations (Walter et al. \cite{walter94}; Preibisch \& Zinnecker \cite{preibisch99}; Haisch et al.\ \cite{haisch01}; Dolan \& Mathieu \cite{dolan99}, \cite{dolan01}). We could interpret our results within the framework of the decreasing disk fraction with age that has been clearly established by Haisch et al.\ (\cite{haisch01} and references therein), using the ($K-L$) excesses found among the low-mass stars of several young clusters and associations. Our results would then support their conclusions, but also extend the range of masses over which the disk decay has been observed to the substellar regime.

In summary, we have found no evidence for a significant population of low-mass stars and brown dwarfs around $\sigma$\,Ori which exhibit K$_{\rm s}$-band excesses. This points strongly towards a low disk frequency, consistent with observations of other similarly aged clusters. Thus this new result adds support to the idea that disk evolution in clusters takes place on short (few Myr) timescales and extends this conclusion to very low masses. We find some evidence to suggest that disk frequencies in the $\sigma$\,Ori association, as measured by the K-band excess, may be mass dependent in the opposite sense to those found in the Trapezium cluster. We do not attach great weight to this result at present because of difficulties in comparing disk fractions measured by different authors using differing techniques but, if true, it might suggest that the disk dissipation timescales were shorter for lower mass objects. More answers may be obtained on the mass dependence of the disk frequency by extending infrared surveys into the brown dwarf domain for more clusters with different ages, and preferably at longer wavelengths, where disk presence can be more reliably diagnosed.


\begin{acknowledgements}
This publication makes use of data products from the Two Micron All Sky Survey, a joint project of the University of Massachusetts and the IPAC, funded by NASA and the National Science Foundation. The 2MASS science data and information services were provided by the InfraRed Science Archive (IRSA) at IPAC, which is operated by JPL. JMO is a PPARC postdoctoral research associate.

\end{acknowledgements}


\end{document}